\documentclass[twocolumn,showpacs,preprintnumbers,superscriptaddress,prl]{revtex4}

\usepackage{amssymb}
\usepackage{graphicx}
\usepackage{epsfig}
\usepackage{amsmath,amssymb}
\usepackage{subfigure}

\begin{document}

\title{Coherent cavity networks with complete connectivity}
\author{Elica S. Kyoseva}
\affiliation{Centre for Quantum Technologies, National University of Singapore, 3 Science Drive 2, Singapore 117543}
\author{Almut Beige}
\affiliation{The School of Physics and Astronomy, University of Leeds, Leeds LS2 9JT, United Kingdom}
\author{Leong Chuan Kwek}
\affiliation{Centre for Quantum Technologies, National University of Singapore, 3 Science Drive 2, Singapore 117543}
\affiliation{National Institute of Education and Institute of Advanced Studies, Nanyang Technological University, 1 Nanyang Walk, Singapore 637616}
\date{\today }

\begin{abstract}
When cavity photons couple to an optical fiber with a continuum of modes, they usually leak out within a finite amount of time. However, if the fiber is about one meter long and linked to a mirror, photons bounce back and forth within the fiber on a much faster time scale. As a result, {\em dynamical decoupling} prevents the cavity photons from entering the fiber. In this paper we use the simultaneous dynamical decoupling of a large number of distant cavities from the fiber modes of linear optics networks to mediate effective cavity-cavity interactions in a huge variety of configurations. Coherent cavity networks with complete connectivity can be created with potential applications in quantum computing and simulation of the complex interaction Hamiltonians of biological systems.
\end{abstract}

\maketitle

\begin{figure}[t]
\begin{minipage}{\columnwidth}
\begin{center}
\includegraphics[scale=0.5]{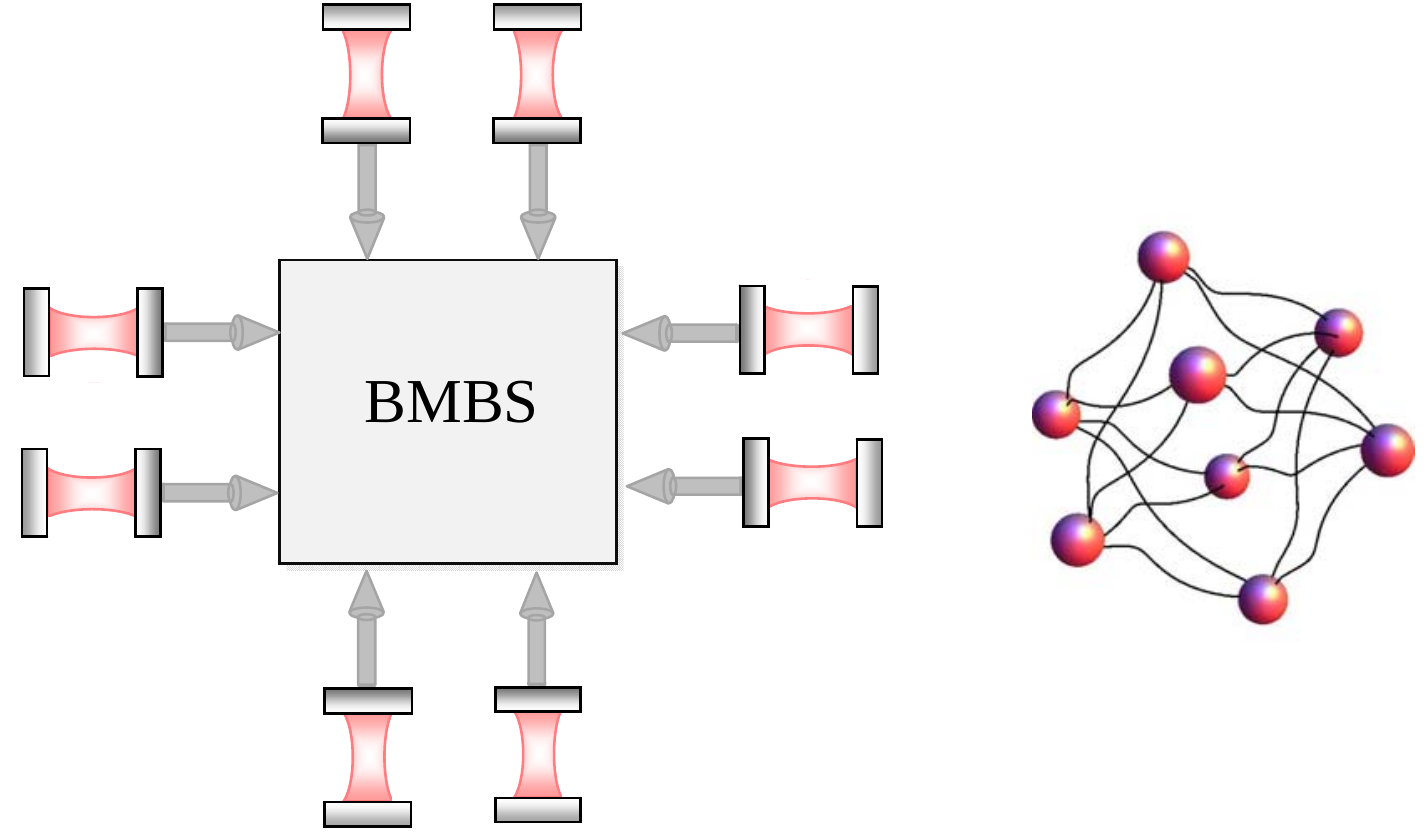}
\end{center} 
\caption{Eight cavities connected through a $4 \times 4$ (BMBS). As illustrated on the right, time scale separation maps the fiber photon couplings of this linear optics network onto effective interactions (represented by lines) between distant cavities (represented by dots) thus creating highly connected cavity networks.
} \label{fig1}
\end{minipage}
\end{figure}

Many applications of quantum information processing, like the simulation of quantum many-body systems, require the construction of scalable qubit networks \cite{lattice}. One of the most prominent examples of such networks are optical lattices with one atom per site and effective interactions induced by Bose-Hubbard-like Hamiltonians \cite{lattice2}. Alternatively one could couple on-demand single photon sources via linear optics elements \cite{Kok,Kwek} or employ the direct coupling of optical cavities via fibers or nanowires \cite{Cirac,Pellizzari2,vanEnk,Bose,Bose2,BuschPRL,Busch}, non-classical light \cite{Kraus,Polzik}, or overlapping evanescent field modes \cite{Plenio,Dimitris}. However, such networks have only effective next neighbour interactions. They are hence not well suited for quantum computing with highly connected graph states \cite{Zanardi} and for simulation of the complex interaction Hamiltonians of biological systems \cite{Vitiello}.  

The purpose of this paper is to show that building scalable cavity networks might be much easier than previously thought. We show how to create coherent cavity architectures with high connectivity in a huge variety of configurations using a dynamical decoupling mechanism \cite{Viola} and coupling optical cavities via optical fibers and linear optics elements, like beam splitters and phase shifters. An example of a highly connected network consisting of eight cavities linked via a Bell multiport beam splitter (BMBS) is shown in Fig.~\ref{fig1}. As we shall see below, it is even possible to achieve {\em complete connectivity}, where each cavity interacts with all other cavities in the network via an effective bi-linear Hamiltonian. 

Suppose the ends of the optical fibers are directly linked to highly reflecting mirrors, namely to the outsides of the mirrors which form the optical cavities. Moreover, in the following we assume that the fibers are about one meter long. In this case, a photon within the fiber would bounce back and forth very rapidly many times before entering the cavity. In other words, the fiber photon modes evolve on a much faster time scale than the cavity photons. This clear separation in the relevant time scales results in the dynamical decoupling of the cavity-fiber transmission from the system dynamics. When several cavities couple simultaneously to the same set of fiber modes, the population of fiber modes is strongly inhibited. However, cavity photons can now tunnel through the linear optics network. The symmetry of the setup mediates effective cavity-cavity interactions. Like the Bose Hubbard-like Hamiltonians of atoms in optical lattices, these can be used to perform quantum computational tasks and to simulate complex quantum phenomena.

To describe experimental setups like the one shown in Fig.~\ref{fig1}, we develop an effective Hamiltonian formalism. Instead of considering the linear optics network as a scattering device, we treat it as a passive symmetric multiport \cite{Stenholm,Stenholm2}. The transmission of photons through the linear optics network at a relatively large rate $J$ and the continuous fiber-cavity transmission with coupling constants $g_{i{\bf k}}$ are modelled by an interaction Hamiltonian. Starting from this Hamiltonian, we account for the different time scales of these two processes via an adiabatic elimination of the fiber photon modes. The result is an effective network Hamiltonian $\mathrm{H}_{\rm eff}$ which maps fiber couplings onto direct cavity-cavity interactions and which is valid for single- and for multi-mode fibers. Notice that our model does not consider the fibers to be high-Q cavities. Our results are consistent with the detailed description of two fiber-linked cavities by van Enk {\em et al.} \cite{vanEnk} but are moreover scalable to complex linear optics networks. Effective cavity interactions are obtained for a much wider regime of experimental parameters than previously assumed. We show that perfect fiber mode matching is not required, since imperfections only result in reduced coupling constants. 

\begin{figure}[t]
\begin{minipage}{\columnwidth}
\centering
\includegraphics[scale=0.6]{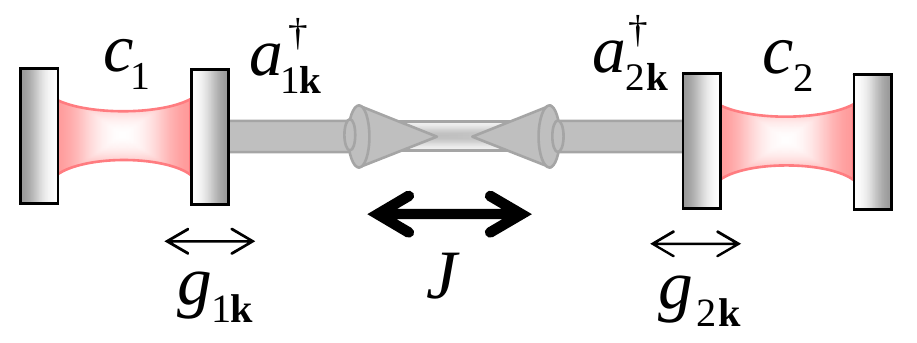} 
\caption{Two fiber coupled optical cavities with the relevant annihilation operators. Cavity photons couple to fiber photons with coupling constants $g_{i {\bf k}}$ while the coupling between different fiber modes is denoted by $J$. As long as Eq.~(\ref{condi}) is valid, dynamical decoupling strongly inhibits the population of fiber photon modes. Nevertheless, due to the symmetries of the system, cavity photons can tunnel through the fiber and induce an effective cavity-cavity interaction (c.f. Eq.~(\ref{Heff1})).}
\label{fig2}
\end{minipage}
\end{figure}

The simplest coherent cavity network consists of only two cavities linked by a fiber as shown in Fig.~\ref{fig2}. This setup has already been discussed in detail in Refs.~\cite{Pellizzari2,vanEnk} using an input-output formalism which describes a photon with a certain initial state entering the fiber at a time $t=0$ and then travelling to the other end. In the following we revisit this example, since it provides much insight into the generalisation to coherent cavity networks proposed here. We show that the setup in Fig.~\ref{fig2} can be modelled much more easily using an effective Hamiltonian formalism which reproduces the results of the input-output formalism for certain initial states. However, it allows us to moreover model the continuous coupling of optical cavities to a fiber connection. Like the model presented in Refs.~\cite{Cirac,Pellizzari2,vanEnk}, we consider multi-mode fibers supporting photons with different wave vectors ${\bf k}$ travelling in different directions and we do not assume that the fiber constitutes a single-mode cavity.

In the following, we denote the bosonic operators which annihilate fiber photons with wave vector ${\bf k}$ and frequency $\omega_k$ traveling to the right and to the left by $a_{1{\bf k}}$ and $a_{2{\bf k}}$, respectively, while $c_i$ annihilates photons in cavity $i$. We assume that both cavities have the same frequency $\omega_{\rm c}$. The interaction Hamiltonian $\mathrm{H}_{\rm I}$ of the two linked cavities with respect to the free energy of the uncoupled systems can then be written as ${\mathrm H}_{\rm I} = {\mathrm H}_{\rm LO} + {\mathrm H}_{\rm LO-cav}$ with
\begin{eqnarray} \label{simple}
{\mathrm H}_{\rm LO} &=& \sum_{\bf k}  \frac{1}{2} \hbar J \, \big[ a_{1 {\bf k}}^\dagger a_{2 {\bf k}} + {\rm H.c.} \big] , \nonumber \\
{\mathrm H}_{\rm LO-cav} &=& \sum_{i=1,2} \sum_{\bf k} \hbar g_{i {\bf k}} \, \mathrm{e}^{\mathrm{i}(\omega_{\textnormal{c}} - \omega_{k})t} \, a_{i{\bf k}}^\dagger c_i + {\rm H.c.} ~~~~
\end{eqnarray}
The Hamiltonian ${\mathrm H}_{\rm LO}$ takes into account that a photon in the $a_{1 {\bf k}}$ mode travels for a finite amount of time $t$ before being reflected by cavity 2. During the reflection, it accumulates a phase factor and becomes a photon of the $a_{2 {\bf k}}$ mode. The result is an effective coupling between different fiber photon modes \cite{Stenholm,Stenholm2}. The corresponding (real) coupling strength $J$ can be calculated easily taking into account that $t = \pi/J = L/ c$, where $L$ is the length of the fiber and $c$ is the speed of light. Since $t$ does not depend on ${\bf k}$ and $L \sim 1\,$m, $J \sim 1 \,$GHz for all photons. 

The second Hamiltonian ${\mathrm H}_{\rm LO-cav}$ in Eq.~(\ref{simple}) models the conversion of fiber into cavity photons due to transmission of the cavity mirrors with $g_{i {\bf k}}$ the corresponding coupling constants. Assuming that the reflection of a fiber photon at a cavity mirror is much more likely than its transmission into the cavity, we can assume that 
\begin{equation} \label{condi}
\sum_{\bf k} g_{i {\bf k}} \ll J, 
\end{equation}
if $J \sim1 \,$GHz. In this case, the Hamiltonian ${\mathrm H}_{\rm I}$ can be simplified via an adiabatic elimination of the fiber modes. The result is the effective network Hamiltonian
\begin{equation} \label{Heff1}
{\rm H}_{\rm eff} = 2 J_{\rm eff} \,  c_1^{\dagger} c_2 + {\rm H.c.} ,
\end{equation}
which describes a direct coupling between the two cavities with the (real) coupling constant 
\begin{equation} \label{effrates}
J_{\rm eff} \equiv - \frac{1}{J} \sum_{\bf k} g_{1 {\bf k}}^* g_{2 {\bf k}} .
\end{equation}
Maximum coupling between the cavities requires a symmetric setup with both cavities coupling equally to the fiber such that $g_{1 {\bf k}} = g_{2 {\bf k}}$. The cavities should be in resonance with each other as much as possible. In the absence of any mode matching, $J_{\rm eff}$ becomes zero. Moreover, Eq.~(\ref{effrates}) shows that the fiber should not be too short, since $J_{\rm eff}$ scales as $c/L$. One way to measure $J_{\rm eff}$ experimentally, is to excite one of the cavities and to observe the time evolution of its photon population.

The above dynamical decoupling mechanism between cavity and fiber photon modes is analogous to the dynamical decoupling of a single qubit from its bosonic bath described by Viola and Lloyd \cite{Viola}. In Ref.~\cite{Viola}, the qubit dynamics becomes inhibited due to strong driving of the bath modes. Here, the leakage of cavity photons into the fiber modes is suppressed as long as the time it takes a photon to travel through the fiber is short compared to the inverse fiber-cavity coupling strength (cf.~Eq.~(\ref{condi})). When coupling two cavities to the same fiber, the dynamical decoupling of the fiber photons nevertheless allows for the tunnelling of cavity photons through the fiber on a time scale proportional to $1/J$ (cf.~Eq.~(\ref{effrates})). Our analysis illustrates that the symmetry of the experimental setup induces a significant cavity-cavity interaction without actually populating the fiber. Note, that the same effective Hamiltonian $\mathrm{H}_{\rm eff}$ applies for experimental setups with single- and with multi-mode fibers. In the following we consider single-mode fibers for simplicity.

\begin{figure}[t]
\centering
\includegraphics[scale=0.5]{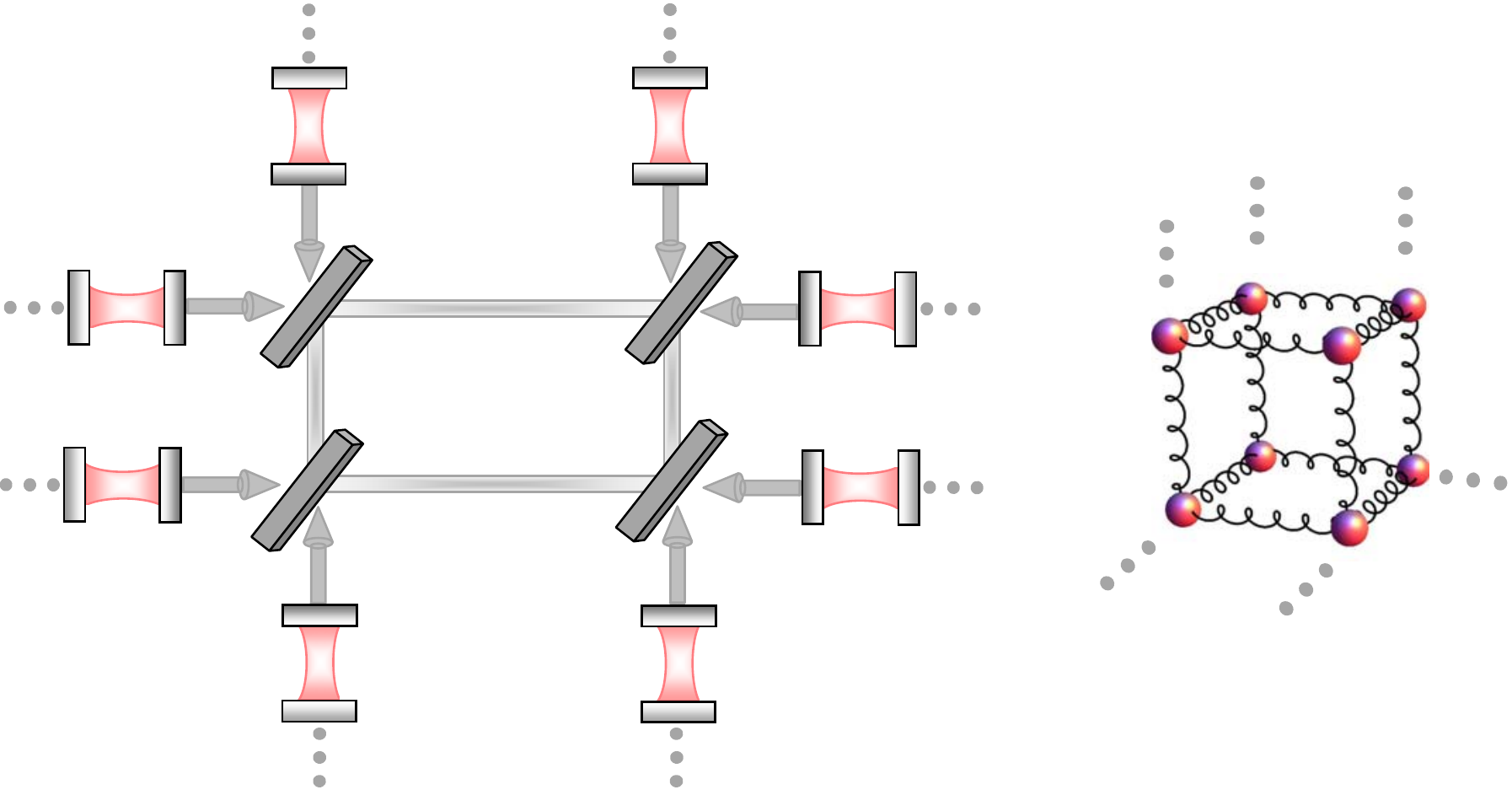} 
\caption{A more complex example of a scalable linear optics network which consists of Mach-Zehnder interferometers which input and output ports contain cavities. As shown on the right, this setup creates an effective 3D cavity network with high connectivity and inherent scalability. The cavities are represented by spheres whereas the effective couplings are represented by spiral lines.} \label{fig3}
\end{figure}

Let us now have a look at what happens when we integrate linear optics elements, like beam splitters and phase shifters, into the fiber. In the following, we describe how to obtain the effective cavity-cavity coupling Hamiltonian $\mathrm{H}_{\rm eff}$ for the case of $2N$ cavities linked via an $N \times N$ linear optics network. The first step is again to derive the time-independent fiber photon conversion Hamiltonian $\mathrm{H}_{\rm LO}$. To do so, we consider photon input states of the form $|\phi_{\textnormal{in}}\rangle = \prod_{i=1}^{2N} \big(\alpha_{i} + \beta_{i} \, a^{\dag}_{i} \big) |0\rangle $ with at most one photon in each fiber mode while ignoring the presence of the cavities. After the photons have passed once through the network, their state can be written as $|\phi_{\textnormal{out}} \rangle = \prod_{i=1}^{2N} \big(\alpha_{i} + \beta_{i} \sum^{2N}_{j=1} U_{ji} \, a^{\dag}_{j} \big) |0\rangle $ \cite{Zukow,Yuan}. The $U_{ij}$ are the matrix elements of a unitary scattering transformation matrix ${\rm U}$ which depends explicitly on the elements of the linear optics network. The Hamiltonian $\mathrm{H}_{\rm LO}$ is then found by demanding that it realises $|\phi_{\textnormal{out}} \rangle = \mathrm{U} \, |\phi_{\textnormal{in}}\rangle$ after a time $t = \pi /J$ \footnote{Here we assume for simplicity that all fibers are approximately of the same length so that all fiber coupling constants $J$ are the same.}. Using the results of  \cite{Stenholm,Stenholm2}, we therefore find that the network interaction Hamiltonian $\mathrm{H}_{\rm I}$ equals $\mathrm{H}_{\rm I} = \mathrm{H}_{\rm LO} + \mathrm{H}_{\rm LO-cav}$ with 
\begin{eqnarray}  \label{stenholm}
\mathrm{H}_{\rm LO} &=& {1 \over 2} \hbar J \, (a_1^\dagger \ldots a_{2N}^\dagger) ({\rm U}-1) (a_1 \ldots a_{2N})^{\rm T} ,\nonumber \\
\mathrm{H}_{\rm LO-cav} &=& \sum_{i=1}^{2N} \hbar g \, \mathrm{e}^{\mathrm{i}(\omega_{\textnormal{c}} - \omega)t} a_i^{\dag} c_{i} + {\rm H.c.}
\end{eqnarray}
The Hamiltonian $\mathrm{H}_{\rm LO-cav}$ accounts again for the transmission of photons through the cavity mirrors. For simplicity, we assume here that all cavities experience the same fiber coupling constant $g$. As long as $g$ is much smaller than $J$, $\mathrm{H}_{\rm I}$ can again be simplified via an adiabatic elimination of the fiber photon modes. The result is an effective network Hamiltonian $\mathrm{H}_{\rm eff}$ which maps the interactions between the different fiber photon modes onto cavity-cavity interactions. Two examples of cavity-coupled $4 \times 4$ linear optics networks are shown in Figs.~\ref{fig1} and \ref{fig3}. In Fig.~\ref{fig1}, the above formalism predicts a direct coupling of each cavity to four other cavities, while the network in Fig.~\ref{fig3} shows a 3D structure which is inherently scalable.

\begin{figure}[t]
\centering
\includegraphics[width=0.8\columnwidth]{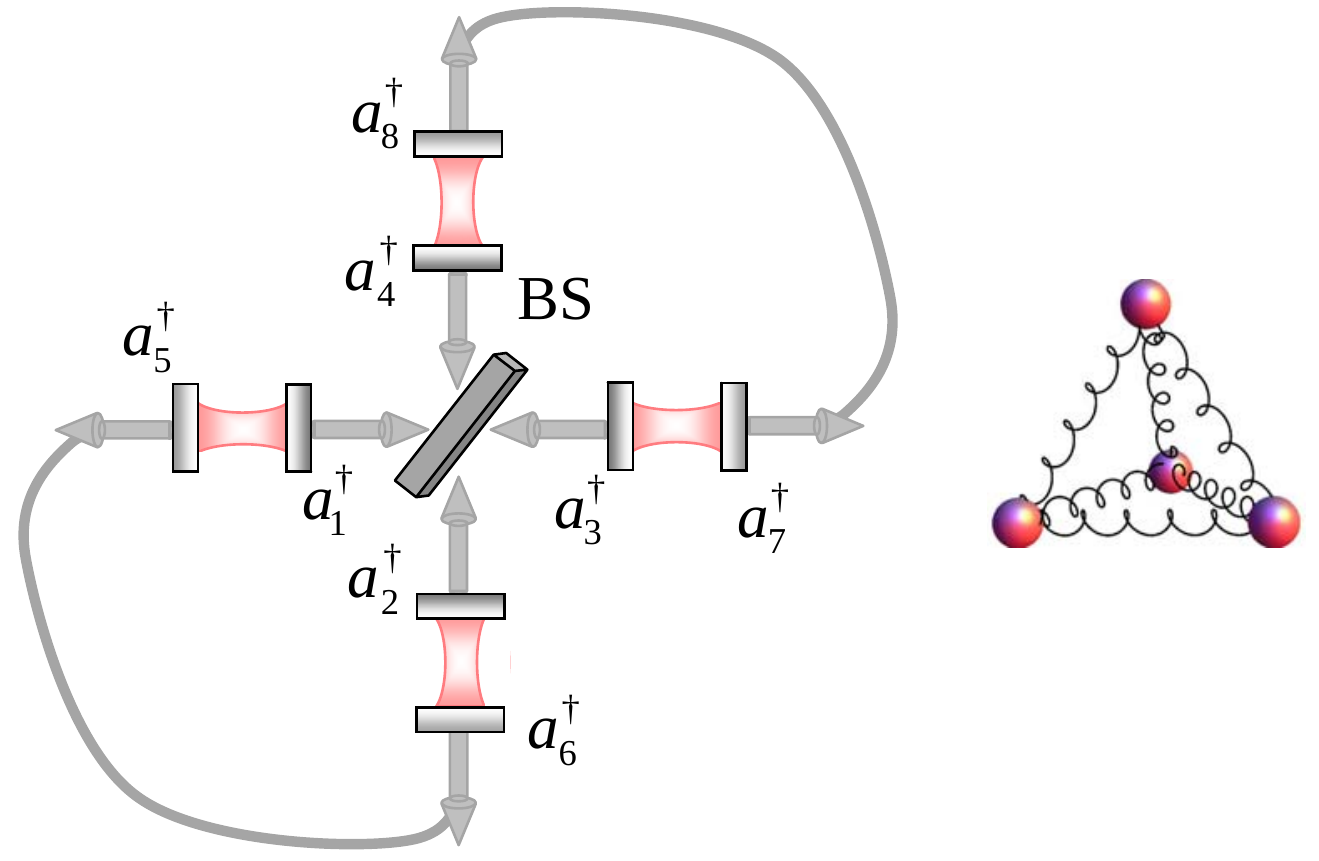} 
\caption{A linear optics network consisting of four cavities, a single beam-splitter (BS), and fiber couplings. Effective interactions are induced between the cavities by fiber photons thus creating a completely connected scalable cavity network in the form of a pyramid.} \label{fig4}
\end{figure}

Using linear optics it is in fact possible to simultaneously connect a single cavity with an arbitrary large number of other cavities. It is even possible to implement any bilinear Hamiltonian of the form $\mathrm{H}_{\rm eff} = \sum_{i,j} \Lambda_{ij} \, c_{i}^\dagger c_j $ by mapping its optical analog $\mathrm{H}_{\rm LO} = \sum_{i,j} \Lambda_{ij} \, a_{i}^\dagger a_j $ \cite{Stenholm,Reck,Ivanov} onto effective network interactions. To illustrate that we now have a closer look at the completely connected cavity network in Fig.~\ref{fig4} which consists of four cavities, one beam-splitter and fiber connections. For this example, the fiber photon conversion Hamiltonian ${\mathrm H}_{\rm LO}$ comprises two fiber couplings as in Eq.~(\ref{simple}) and one network coupling as in Eq.~(\ref{stenholm}) where ${\rm U}$ is the unitary $4 \times 4$ beam-splitter transformation matrix with $U_{13}=U_{24}=U_{31}=U_{42}={1} / \sqrt{2}$, $U_{32}=U_{41}={\rm i} / \sqrt{2}$, $U_{14}=U_{23}=-{\rm i /\sqrt{2}}$, and all other matrix elements equal to zero \cite{Zukow,Yuan}. The interaction Hamiltonian $\mathrm{H}_{\rm I} = \mathrm{H}_{\rm LO} + \mathrm{H}_{\rm LO-cav}$ is hence given by
\begin{eqnarray} \label{TMH}
\mathrm{H}_{\rm LO} &=& \frac{\hbar J}{2\sqrt{2}} \, \big[ a_{1}^\dagger ( a_{3} - {\rm i} 
a_{4} \big) - {\rm i} a_{2}^\dagger ( a_{3 } + {\rm i} a_{4} \big) + a_{5}^\dagger a_{6} \nonumber \\ 
&& + a_{7}^\dagger a_{8} + {\rm H.c.} \big] , \nonumber \\
\mathrm{H}_{\rm LO-cav} & = & \sum_{i=1}^4 \hbar g \, \mathrm{e}^{\mathrm{i}(\omega_{\textnormal{c}} - \omega_{\rm f})t} \, a^{\dag}_i c_{i } + {\rm H.c.}
\end{eqnarray}
Proceeding as before and adiabatically eliminating the fiber photons, we now obtain the effective network Hamiltonian
\begin{eqnarray} \label{Heff}
\mathrm{H}_{\rm eff} &=& \sqrt{2} \hbar J_{\rm eff} \, \big[c_1^\dagger (c_2 + c_3 - {\rm i} c_4 \big) - {\rm i} c_2^\dagger (c_1 + c_3 + {\rm i} c_4 \big) \nonumber \\
&& + c_3^\dagger c_4 + {\rm H.c.} \big]
\end{eqnarray}
with $J_{\rm eff} = - |g|^2/J$. As illustrated in Fig.~\ref{fig4}, each cavity is linked to all other cavities. A completely connected network of four cavities is created. By connecting more cavities via fiber based linear optics networks, even larger and more complex networks with high and complete connectivity can be created.

{\em In conclusion}, we have shown that it is possible to create coherent cavity networks with complete connectivity using only fiber couplings and linear optics elements. To show how this works we first analysed a network which contains only two cavities connected by a single multi-mode fiber. When the cavities have a much higher reflectivity than transmission coefficient on the inside \emph{and} on the outside of their mirrors, a fast time scale is introduced which presence suppresses the relatively slow leakage of cavity photons into the fiber. Nevertheless, effective cavity-cavity interactions are created when several cavities couple to a set of common fiber photon modes. In this paper we illustrated this coupling mechanism for distant cavities with concrete examples.

Our analysis obtains the same effective network Hamiltonian $\mathrm{H}_{\rm eff} $ for multi- and for single-mode fibers. Moreover, it shows that the fiber can be relatively long (about $1\,$m) as long as there is a clear separation between the time it takes a photon to travel once through the linear optics setup and the average time it takes a photon to leak out of a cavity (c.f.~Eq.~(\ref{condi})). Let us also point out that we require neither interferometric stability nor perfect mode matching and can tolerate relatively larger fiber photon loss rates. The reason for this is that our analysis eliminates all fiber photon modes adiabatically from the system dynamics, i.e., there are on average almost no photons in the fiber. Fiber photon modes ${\bf k}$ without mode matching (i.e.~$g_{i{\bf k}}^* g_{j{\bf k}} = 0$) do not contribute to $\mathrm{H}_{\rm eff} $. As long as there is a sufficient overlap between the relevant subsets of fiber photon modes and as long as the cavities are more or less in resonance, we expect relatively strong cavity-cavity interactions with the coupling constant $J_{\rm eff}$ given in Eq.~(\ref{effrates}). Slight variations of the fiber length cause only small changes of the amplitude of $J_{\rm eff}$ but introduce phase fluctuations due to variations of the $g_{i {\bf k}}$. However, it is not important that the length of the fiber is a multiple of the cavity wave length $\lambda_{\rm c}$, since the fibers do not need to resemble cavities.

The effective Hamiltonian formalism introduced here does not account for the loss of photons, for example, due to absorption within the cavity mirrors. Such losses can be included using the usual master equation approach which assigns spontaneous decay rates to the cavities. As long as these decay rates remain small compared to $J_{\rm eff}$, the above proposed network architecture is expected to find many applications in quantum information processing. Such applications might require the coupling of single laser-driven atoms or quantum dots to the cavities. However, the effective network Hamiltonian $\mathrm{H}_{\rm eff} $ remains valid as long as the time scale given by $1/J$ for the time evolution of the fiber photon modes is the fastest. Using fiber-based linear optics networks, a huge variety of coherent cavity networks can be created. Due to the high versatility of linear optics \cite{Stenholm,Reck,Ivanov}, it is possible to couple cavities to \emph{any} number of other cavities. They are hence well suited for quantum computing with highly connected graph states \cite{Zanardi} and for simulation of the complex interaction Hamiltonians of biological systems \cite{Vitiello}.  \\[0.2cm]
{\em Acknowledgement.} A.B.~acknowledges stimulating discussions with J. Busch and a James Ellis University Research Fellowship from the Royal Society and the GCHQ. E.S.K. and L.C.K. acknowledge financial support by the National Research Foundation and the Ministry of Education, Singapore.

\end{document}